\documentclass[aps,prd,nofootinbib,amsmath,amssymb,superscriptaddress,twocolumn,10pt]{revtex4}%superscriptaddress,showpacs,

\usepackage{graphicx}
\usepackage{dcolumn}
\usepackage{bm}
\usepackage{amssymb}
\usepackage{latexsym}
\usepackage{booktabs}
\usepackage{amsmath}
\usepackage{multirow}
\usepackage[colorlinks=true, linkcolor=red, citecolor=blue]{hyperref}

\def\be{\begin{equation}}
\def\ee{\end{equation}}
\def\ba{\begin{eqnarray}}
\def\ea{\end{eqnarray}}

\begin{document}

\title{Search for sterile neutrinos in a universe of vacuum energy interacting with cold dark matter}

\author{Lu Feng}
%\email{fengluu@foxmail.com}
\affiliation{Department of Physics, College of Sciences, Northeastern University, Shenyang
110004, China}
\author{Jing-Fei Zhang}
%\email{jfzhang@mail.neu.edu.cn}
\affiliation{Department of Physics, College of Sciences, Northeastern University, Shenyang
110004, China}
\author{Xin Zhang}%\footnote{Corresponding author}}
%\email{zhangxin@mail.neu.edu.cn}
\affiliation{Department of Physics, College of Sciences, Northeastern University, Shenyang 110004, China}
\affiliation{Center for High Energy Physics, Peking University, Beijing 100080, China}

\begin{abstract}
We investigate the cosmological constraints on sterile neutrinos in a universe in which vacuum energy interacts with cold dark matter by using latest observational data. We focus on two specific interaction models, $Q=\beta H\rho_{\rm v}$ and $Q=\beta H\rho_{\rm c}$. To overcome the problem of large-scale instability in the interacting dark energy scenario, we employ the parametrized post-Friedmann (PPF) approach for interacting dark energy to do the calculation of perturbation evolution. The observational data sets used in this work include the Planck 2015 temperature and polarization data, the baryon acoustic oscillation measurements, the type-Ia supernova data, the Hubble constant direct measurement, the galaxy weak lensing data, the redshift space distortion data, and the Planck lensing data. Using the all-data combination, we obtain $N_{\rm eff}<3.522$ and $m_{\nu,{\rm sterile}}^{\rm eff}<0.576$ eV for the $Q=\beta H\rho_{\rm v}$ model, and $N_{\rm eff}=3.204^{+0.049}_{-0.135}$ and $m_{\nu,{\rm sterile}}^{\rm eff}=0.410^{+0.150}_{-0.330}$ eV for the $Q=\beta H\rho_{\rm c}$ model. The latter indicates $\Delta N_{\rm eff}>0$ at the 1.17$\sigma$ level and a nonzero mass of sterile neutrino at the 1.24$\sigma$ level. In addition, for the $Q=\beta H\rho_{\rm v}$ model, we find that $\beta=0$ is consistent with the current data, and for the $Q=\beta H\rho_{\rm c}$ model, we find that $\beta>0$ is obtained at more than 1$\sigma$ level.
\end{abstract}
%\pacs{95.36.+x, 98.80.Es, 98.80.-k}
\maketitle

\section{Introduction}
\label{sec1}
At its present stage, our universe is undergoing an accelerating expansion, which has been confirmed by a number of astronomical observations, such as type Ia supernovae \cite{Riess:1998cb,Perlmutter:1998np}, cosmic microwave background \cite{Jarosik:2006ib,Page:2006hz,Spergel:2006hy,Spergel:2003cb,Bennett:2003bz}, and large-scale structure \cite{Tegmark:2006az,Tegmark:2003uf,Tegmark:2003ud}. The acceleration of the universe strongly indicates the existence of ``dark energy" \cite{Li:2012dt,Bamba:2012cp,Li:2011sd,Kamionkowski:2007wv,Sahni:2006pa,Bean:2005ru,Peebles:2002gy} with negative pressure. The preferred candidate for dark energy is the cosmological constant $\Lambda$ proposed by Einstein \cite{Einstein:1917ce}. The equation-of-state parameter of $\Lambda$ is $w_{\Lambda}\equiv p_{\Lambda}/\rho_{\Lambda}=-1$. The cosmological model with $\Lambda$ and cold dark matter (CDM) is usually called the $\Lambda$CDM model, which fits observations quite well. However, the cosmological constant is plagued with several theoretical difficulties, such as the so-called ``fine-tuning" and ``cosmic coincidence" problems \cite{Weinberg:2000yb,Carroll:2000fy}. These theoretical puzzles have led to significant efforts in developing alternative scenarios, such as dynamical dark energy models or modified gravity theories. In particular, the interacting dark energy scenario has been proven by cosmologists to be capable of alleviating the coincidence problem effectively \cite{Comelli:2003cv,Cai:2004dk,Zhang:2005rg,He:2008tn,He:2009pd}.

The interacting dark energy (IDE) scenario, in which it is considered that there is some direct, non-gravitational coupling between dark energy and dark matter, has been widely studied \cite{Guo:2017deu,Amendola:1999qq,Amendola:1999er,TocchiniValentini:2001ty,Amendola:2001rc,Comelli:2003cv,Chimento:2003iea,Cai:2004dk,Zhang:2005rj,Zhang:2005rg,Zimdahl:2005bk,Wang:2006qw,Sadjadi:2006qp,Guo:2007zk,Bertolami:2007zm,Zhang:2007uh,Abdalla:2007rd,Szydlowski:2008by,Boehmer:2008av,He:2008tn,Bean:2008ac,Chen:2008ft,Gavela:2009cy,He:2009mz,CalderaCabral:2009ja,He:2009pd,Koyama:2009gd,Majerotto:2009np,Valiviita:2009nu,Li:2009zs,Zhang:2009qa,Boehmer:2009tk,Xia:2009zzb,Chimento:2009hj,Cai:2009ht,He:2010ta,Martinelli:2010rt,Gavela:2010tm,Cui:2010dr,Wei:2010cs,Li:2010eu,Li:2010ak,He:2010im,Xu:2011tsa,Li:2011ga,Clemson:2011an,Fu:2011ab,Zhang:2012uu,Salvatelli:2013wra,Zhang:2013lea,Xu:2013jma,Zhang:2013zyn,Li:2013bya,Yang:2014gza,yang:2014vza,Wang:2014iua,Faraoni:2014vra,Wang:2014oga,Salvatelli:2014zta,Geng:2015ara,Duniya:2015nva,Yin:2015pqa,Cai:2015zoa,Li:2015vla,Murgia:2016ccp,Wang:2016lxa,Pourtsidou:2016ico,Costa:2016tpb,Sola:2016ecz,Feng:2016djj,Xia:2016vnp,vandeBruck:2016hpz,Sola:2016zeg,Kumar:2017dnp,Guo:2017hea,Zhang:2017ize,Li:2017usw}. The IDE scenario can provide more features to fit the observations, because it not only can affect the expansion history of the universe, but also can change the growth history in a more direct way. The impacts on the cosmic microwave background (CMB) \cite{He:2009pd,Pourtsidou:2016ico} and the large-scale structure (LSS) formation \cite{Amendola:2001rc,Bertolami:2007zm,He:2009mz,Koyama:2009gd,Li:2013bya,Pourtsidou:2016ico} in the IDE scenario have been discussed in detail.

Recently, Ref.~\cite{Guo:2017hea} explores the cosmological weighing of active neutrinos in the scenario of vacuum energy interacting with cold dark matter by using current cosmological observations. It is found that, when the interaction is considered, the upper limit of active neutrino mass may be largely changed. That is to say, the interaction between vacuum energy and cold dark matter may affect the cosmological weighing of active neutrinos. In this work, we wish to investigate how the interaction between vacuum energy and cold dark matter would affect the constraints on sterile neutrino parameters with the  current cosmological observations.

The existence of light massive sterile neutrinos is hinted by the anomalies of short-baseline (SBL) neutrino oscillation experiments \cite{Mention:2011rk,Conrad:2012qt,Giunti:2012tn,Kopp:2013vaa,Giunti:2013aea,Gariazzo:2013gua}. The SBL neutrino oscillation experiments seem to require the existence of an extra sterile neutrino with mass at the eV-scale. However, it seems that this result is not favored by some recent studies, such as the neutrino oscillation experiment by the Daya Bay and MINOS collaborations~\cite{Adamson:2016jku}, the cosmic-ray (atmospheric neutrino) experiment by the IceCube collaboration~\cite{TheIceCube:2016oqi}, and the latest cosmological searches~\cite{Zhang:2014dxk,Feng:2017nss,Zhao:2017urm,Feng:2017mfs}. See Refs.~\cite{Zhao:2017jma,Li:2014dja,Zhang:2014lfa,Zhang:2014nta,Zhang:2014ifa,Dvorkin:2014lea,Wyman:2013lza,Hamann:2013iba,Battye:2013xqa} for more recent relevant studies.

In this paper, we investigate the search for sterile neutrinos in a scenario of vacuum energy interacting with cold dark matter (IvCDM, hereafter). The purpose of considering the vacuum energy ($w=-1$) is to investigate the pure impact of the coupling on the constraints on sterile neutrino parameters.

In the IvCDM model, the energy conservation equations for the vacuum energy and cold dark matter satisfy
\begin{align}
&\rho'_{\rm v} = aQ,\label{eq1}\\
&\rho'_{\rm c} = -3\mathcal{H}\rho_{\rm c}-aQ,\label{eq2}
\end{align}
where $\mathcal{H}$ is the conformal Hubble parameter defined as $\mathcal{H}=a'/a$ ($a$ is the scale factor of the universe), a prime denotes the derivative with respect to the conformal time $\eta$, $\rho_{\rm v}$ and $\rho_{\rm c}$ represent the energy densities of vacuum energy and cold dark matter, respectively, and $Q$ denotes the energy transfer rate. %For the IDE model, several forms for $Q$ have been put forward \cite{Amendola:1999er,Cai:2004dk,Zhang:2005rg,Guo:2007zk,Boehmer:2008av,He:2009pd,Xia:2009zzb,Li:2010ak,Li:2011ga,Zhang:2012uu,Li:2013bya,Wang:2014oga,Feng:2016djj,Li:2017usw}.
In this work, we consider two popular forms of $Q$, i.e., $Q=\beta H\rho_{\rm v}$ (denoted as $Q_1$) and $Q=\beta H\rho_{\rm c}$ (denoted as $Q_2$), where $\beta$ is the dimensionless coupling constant and $H$ is the Hubble expansion rate of the universe, $H=\dot{a}/a$, where the dot denotes the derivative with respect to the cosmic time $t$. From Eqs.~(\ref{eq1}) and (\ref{eq2}), it is indicated that $\beta<0$ means that vacuum energy decays into cold dark matter, $\beta>0$ means that cold dark matter decays into vacuum energy, and $\beta=0$ corresponds to the case without interaction. For convenience, the IvCDM models with $Q_1$ and $Q_2$ are also denoted as the IvCDM1 model and the IvCDM2 model, respectively.

This paper is organized as follows. In Sec. \ref{sec2}, we introduce the method and the observational data used in this paper. In Sec. \ref{sec3}, we present the fit results and discuss these results in detail. Conclusion is given in Sec. \ref{sec4}.

\section{Method and Data}
\label{sec2}

\subsection{Method}

%In this paper, our aim is to search for sterile neutrinos in the IvCDM model by using the latest observation data.

For the IvCDM model, the base parameter set (including seven free parameters) is \{$\Omega_bh^2$, $\Omega_ch^2$, $100\theta_\ast$, $\beta$, $\tau$, $\ln (10^{10}A_s)$, $n_s$\}, where $\Omega_bh^2$ is the physical baryon density, $\Omega_ch^2$ is the physical cold dark matter density, $\theta_\ast$ is the ratio between the sound horizon and the angular diameter distance at the time of last-scattering, $\beta$ is the coupling constant characterizing the interaction strength between vacuum energy and cold dark matter, $\tau$ is the Thomson scattering optical depth due to reionization, $A_s$ is the amplitude of the primordial curvature perturbation, and $n_s$ is its power-law spectral index.

When massive sterile neutrinos are considered in the IvCDM model, two extra free parameters, the effective number of relativistic species $N_{\rm eff}$ and the effective sterile neutrino mass $m_{\nu,{\rm sterile}}^{\rm eff}$, should be involved in the calculation. The true mass of a thermally-distributed sterile neutrino ($m_{{\rm sterile}}^{\rm thermal}$) is related to these two parameters by $m_{\nu,{\rm sterile}}^{\rm eff}=(N_{\rm eff}-3.046)^{3/4}m_{{\rm sterile}}^{\rm thermal}$. Here, in order to avoid a negative $m_{{\rm sterile}}^{\rm thermal}$, $N_{\rm eff}$ must be larger than 3.046. In this case, the total mass of active neutrinos $\sum m_\nu$ is fixed at 0.06 eV.
Since we add massive sterile neutrinos into the IvCDM model, the model considered in this paper is called the IvCDM+$\nu_s$ model. For convenience, in this paper, we use IvCDM1+$\nu_s$ and  IvCDM2+$\nu_s$ to denote the corresponding $Q=\beta H\rho_{\rm v}$ and $Q=\beta H\rho_{\rm c}$ models that involve sterile neutrinos, respectively.
Thus, there are nine base parameters in the  IvCDM+$\nu_s$ (IvCDM1+$\nu_s$ and  IvCDM2+$\nu_s$) models.

In this work, we also consider the cases with massive active neutrinos, for a comparison. Namely, we will also let $\sum m_\nu$ be a free parameter in the corresponding models. For simplicity, in the cases of active neutrinos, we use IvCDM1+$\nu_a$ and IvCDM2+$\nu_a$ to denote the corresponding $Q=\beta H\rho_{\rm v}$ and $Q=\beta H\rho_{\rm c}$ models. In the cases of considering both active and sterile neutrinos, we use $\Lambda$CDM+$\nu_s$+$\nu_a$, IvCDM1+$\nu_s$+$\nu_a$, and IvCDM2+$\nu_s$+$\nu_a$ to denote the corresponding $Q=0$, $Q=\beta H\rho_{\rm v}$, and $Q=\beta H\rho_{\rm c}$ models.

Note that, when we consider the situation of vacuum energy interacting with cold dark matter, the vacuum energy would no longer be a pure background and in this case we must consider the perturbations of the vacuum energy.
There are large-scale instabilities in the calculations of dark energy perturbations in a conventional way, due to the incorrect calculation of the pressure perturbation of dark energy.
To overcome this difficulty, we employ the parametrized post-Friedmann (PPF) scheme to solve the instability problem of the IDE cosmology \cite{Li:2014eha,Li:2014cee,Li:2015vla,Guo:2017hea,Zhang:2017ize,Feng:2018yew}.
The main idea of the PPF is to establish a direct relationship between the velocities of dark energy and other components on large scales. On small scales, the evolution of curvature perturbation is described by the Possion equation. In order to make these two limits compatible, a dynamical function $\Gamma$ is introduced to link them. By using the Einstein equations and the conservation equations, the equation of motion of $\Gamma$ on all scales can be obtained. Then, we can get energy density and velocity perturbations of dark energy. The PPF framework can give stable cosmological perturbations for the IDE scenario and help us to probe the whole parameter space of the model including $\beta$. For more detailed information about the PPF framework for the IDE models, see Refs. \cite{Li:2014eha,Li:2014cee,Feng:2018yew}.

%we can get the equation of motion of $\Gamma$ on all scale
%\begin{equation}
%(1 + c_\Gamma^2 k_H^2) [\Gamma' +\mathcal{H} \Gamma + c_\Gamma^2 k_H^2 \mathcal{H}\Gamma] = S.
%\label{eqn:gammaeom}
%\end{equation}

%Then, we can get energy density and velocity perturbations of dark energy
%\begin{align}
%&\rho_{\rm de}\Delta_{\rm de} =- 3(\rho_{\rm de}+p_{\rm de}) {V_{\rm de}-V_{T}\over k_{H} }-{k^{2}\ck \over 4\pi G a^{2}} \Gamma,\\ \label{eqn:ppffluid}
%& V_{\rm de}-V_{T} ={-k \over 4\pi Ga^2 (\rho_{\rm de} + p_{\rm de}) F} \nonumber \\
%&\quad\quad\quad\times\left[ S - \Gamma' - \mathcal{H}\Gamma + f_{\zeta}{4\pi Ga^2 (\rho_{T}+p_{T}) \over k}V_{T}
%\right],
%\end{align}
%with $F = 1 +  12 \pi G a^2 (\rho_T + p_T)/( k^2 \ck)$.

Our calculations are based on the public Markov-chain Monte Carlo package {\tt CosmoMC}~\cite{Lewis:2002ah}. To solve the background and perturbation equations for the IvCDM models, we employ the modified PPF code~\cite{Li:2014eha,Li:2014cee,Li:2015vla} (for the original one, see Refs.~\cite{Hu:2008zd,Fang:2008sn}) and make some changes for the Boltzmann code {\tt CAMB}~\cite{Lewis:1999bs}.

\subsection{Observational data}

For observational data, we employ the Planck data in combination with other cosmological probe data. For the Planck data, we use the CMB temperature and polarization anisotropies data, including lowP, TT, TE, and EE, which were released by the Planck collaboration \cite{Aghanim:2015xee}. We shall use ``Planck" to denote the above data combination. For other observations, we consider the following data sets:
\begin{itemize}
\item  BAO: We employ the BAO data including the measurements from 6dFGS ($z_{\rm eff}=0.1$) \cite{Beutler:2011hx}, SDSS-MGS ($z_{\rm eff}=0.15$) \cite{Ross:2014qpa}, and LOWZ ($z_{\rm eff}=0.32$) and CMASS ($z_{\rm eff}=0.57$) samples of BOSS DR12 \cite{Cuesta:2015mqa}.

\item  SN: We use the Joint Light-curve Analysis (JLA) sample \cite{Betoule:2014frx} of type Ia supernovae data, complied from the SNLS, SDSS, and several samples of low-redshift light-curve analysis.

\item  $H_0$: We employ the latest direct measurement of the Hubble constant, $H_0=73.24\pm1.74$ km~s$^{-1}~$Mpc$^{-1}$ \cite{Riess:2016jrr}.

\item  WL: We use the cosmic shear measurement of the galaxy weak lensing from the CFHTLens Survey \cite{Heymans:2013fya}.

\item  RSD: We use the two latest redshift space distortions data from the LOWZ ($z_{\rm eff}=0.32$) sample and the CMASS ($ z_{\rm eff}=0.57$) sample of BOSS DR12 \cite{Gil-Marin:2016wya}.

\item  lensing: We employ the CMB lensing power spectrum from the Planck lensing measurement \cite{Ade:2015zua}.
\end{itemize}

For simplicity, we will use ``BSH" to denote the joint BAO+SN+$H_0$ data combination, and use ``LSS" to denote the joint WL+RSD+lensing data combination.
Thus, in our analysis, we use the two data combinations: (1) Planck+BSH and (2) Planck+BSH+LSS. We will report the constraint results in the next section.

Note that in this work we try to use the observational data sets that have also been used in the fit analyses of cosmological parameters performed by the Planck collaboration \cite{Ade:2015xua,Ade:2015rim,Ade:2015lrj}. Actually, there are also some other additional data sets, such as KiDS, Planck SZ cluster counts, $H(z)$ measurements, and strong gravitational lensing, etc., but we do not use these data in this work, for the following reasons: (1) We wish to be in accordance with the Planck collaboration in the use of data in constraining cosmological parameters, in order for the reader to easily make a comparison. For the weak gravitational lensing observation, we thus choose to use the CFHTLenS data but not the KiDS data. (2) There are still some uncontrolled systematics and calibration problems in some observations, such as the Planck SZ cluster counts measurements. (3) For observations measuring the expansion history of the universe, the ``BSH'' combination used in this work can provide rather tight constraints on parameters of various dark energy models, but the $H(z)$ and strong lensing observations can only provide fairly weak additional constraints. In addition, we also mention that the primordial light elements (helium and deuterium) abundances predicted by the standard big bang nucleosynthesis (BBN) can also be used to provide constraints on $N_{\rm eff}$. Combination with BBN in the data is expected to provide a somewhat tighter constraint on $N_{\rm eff}$ (see, e.g., Refs.~\cite{Ade:2015xua,Zhang:2014dxk}), but we do not use the BBN data in this work, which may be left to a future further work on this aspect.

%%table%%%%%%%%%%%%%%%%%%%%%%%%%%%%%%%%%%%%%%%%%%%%%%%%%%%%%%%

\begin{table*}\small
\setlength\tabcolsep{0.7pt}
\renewcommand{\arraystretch}{1.5}
\centering
\caption{\label{tabvs}Fit results for the $\Lambda$CDM+$\nu_s$, IvCDM1 ($Q=\beta H\rho_{\rm v}$)+$\nu_s$, and IvCDM2 ($Q=\beta H\rho_c$)+$\nu_s$ models by using Planck+BSH and Planck+BSH+LSS.}
\begin{tabular}{ccccccccc}

\hline Model &\multicolumn{2}{c}{$\Lambda$CDM+$\nu_s$ ($Q=0$)}&&\multicolumn{2}{c}{IvCDM1+$\nu_s$ ($Q=\beta H\rho_{\rm v}$)}&&\multicolumn{2}{c}{IvCDM2+$\nu_s$ ($Q=\beta H\rho_{\rm c}$)}\\
           \cline{2-3}\cline{5-6}\cline{8-9}
       Data  & Planck+BSH &Planck+BSH+LSS &&Planck+BSH &Planck+BSH+LSS &&Planck+BSH &Planck+BSH+LSS\\

\hline
$\Omega_{\rm m}$                         &$0.302\pm0.007$
                                         &$0.305\pm0.007$
                                         &&$0.333^{+0.069}_{-0.053}$
                                         &$0.310^{+0.039}_{-0.031}$
                                         &&$0.296\pm0.008$
                                         &$0.297\pm0.008$\\

$\sigma_8$                               &$0.836^{+0.024}_{-0.019}$
                                         &$0.798^{+0.023}_{-0.016}$
                                         &&$0.818^{+0.040}_{-0.039}$
                                         &$0.793\pm0.021$
                                         &&$0.821^{+0.037}_{-0.024}$
                                         &$0.785^{+0.026}_{-0.022}$\\

$H_0\,[{\rm km}/{\rm s}/{\rm Mpc}]$      &$69.50^{+1.00}_{-1.20}$
                                         &$68.80^{+0.61}_{-1.03}$
                                         &&$69.50\pm1.10$
                                         &$68.84^{+0.70}_{-1.09}$
                                         &&$69.50^{+0.70}_{-0.92}$
                                         &$69.31\pm0.70$\\

$\beta$                                  &...
                                         &...
                                         &&$-0.13^{+0.22}_{-0.30}$
                                         &$-0.02^{+0.14}_{-0.18}$
                                         &&$0.0022\pm0.0015$
                                         &$0.0026\pm0.0015$\\

$N_{\rm eff}$                            &$<3.619$
                                         &$<3.511$
                                         &&$<3.641$
                                         &$<3.522$
                                         &&$<3.498$
                                         &$3.204^{+0.049}_{-0.135}$\\

$m_{\nu,{\rm{sterile}}}^{\rm{eff}}\,[\rm eV]$     &$<0.285$
                                                  &$<0.499$
                                                  &&$<0.312$
                                                  &$<0.576$
                                                  &&$<0.875$
                                                  &$0.410^{+0.150}_{-0.330}$\\

%$m_{{\rm sterile}}^{\rm thermal}\,[\rm eV]$       &$...$
%                                                  &$...$
 %                                                 &&$...$
 %                                                 &$...$
%                                                  &&$...$
%                                                  &$1.636$\\
%\hline
%$\chi^2_{\rm min}$                     &13666.154
%                                       &13708.702
%                                       &&13665.030
 %                                      &13707.830
 %                                      &&13665.384
%                                       &13706.390\\
\hline
\end{tabular}

\end{table*}

%%%%%%%%%%%%%%%%%%%%%%%%%%%%%%%%%%%%%%%%%%%%%%%%%%%%%%%%%%%%%%%%

%%table%%%%%%%%%%%%%%%%%%%%%%%%%%%%%%%%%%%%%%%%%%%%%%%%%%%%%%%

\begin{table*}\small
\setlength\tabcolsep{0.7pt}
\renewcommand{\arraystretch}{1.5}
\centering
\caption{\label{tabvsva}Fit results for the $\Lambda$CDM+$\nu_s$+$\nu_a$, IvCDM1 ($Q=\beta H\rho_{\rm v}$)+$\nu_s$+$\nu_a$, and IvCDM2 ($Q=\beta H\rho_c$)+$\nu_s$+$\nu_a$ models by using Planck+BSH and Planck+BSH+LSS.}
\begin{tabular}{ccccccccc}

%\hline
\hline Model &\multicolumn{2}{c}{$\Lambda$CDM+$\nu_s$+$\nu_a$ ($Q=0$)}&&\multicolumn{2}{c}{IvCDM1+$\nu_s$+$\nu_a$ ($Q=\beta H\rho_{\rm v}$)}&&\multicolumn{2}{c}{IvCDM2+$\nu_s$+$\nu_a$ ($Q=\beta H\rho_{\rm c}$)}\\
           \cline{2-3}\cline{5-6}\cline{8-9}
       Data  & Planck+BSH &Planck+BSH+LSS &&Planck+BSH &Planck+BSH+LSS &&Planck+BSH &Planck+BSH+LSS\\

\hline
$\Omega_m$                                   &$0.302\pm0.007$
                                             &$0.305\pm0.007$
                                             &&$0.329^{+0.075}_{-0.056}$
                                             &$0.309^{+0.038}_{-0.030}$
                                             &&$0.296\pm0.008$
                                             &$0.297\pm0.008$\\

$\sigma_8$                                   &$0.837^{+0.026}_{-0.019}$
                                             &$0.797^{+0.023}_{-0.016}$
                                             &&$0.822^{+0.040}_{-0.044}$
                                             &$0.795\pm0.021$
                                             &&$0.821^{+0.038}_{-0.025}$
                                             &$0.783^{+0.026}_{-0.022}$\\

$H_0\,[{\rm km}/{\rm s}/{\rm Mpc}]$          &$69.53^{+0.93}_{-1.23}$
                                             &$68.85^{+0.68}_{-1.02}$
                                             &&$69.60^{+1.00}_{-1.20}$
                                             &$68.91^{+0.76}_{-1.08}$
                                             &&$69.50^{+0.72}_{-0.91}$
                                             &$69.25^{+0.72}_{-0.73}$\\

$\beta$                                      &...
                                             &...
                                             &&$-0.12^{+0.23}_{-0.32}$
                                             &$-0.02^{+0.13}_{-0.17}$
                                             &&$0.0022\pm0.0016$
                                             &$0.0028\pm0.0016$\\

$\sum m_\nu$\,[\rm eV]                       &$<0.126$
                                             &$<0.173$
                                             &&$<0.131$
                                             &$<0.158$
                                             &&$<0.179$
                                             &$<0.246$\\

$N_{\rm{eff}}$                               &$<3.612$
                                             &$<3.533$
                                             &&$<3.654$
                                             &$<3.537$
                                             &&$<3.499$
                                             &$3.199^{+0.047}_{-0.136}$\\

$m_{\nu,{\rm{sterile}}}^{\rm{eff}}$\,[\rm eV]          &$<0.343$
                                                       &$<0.563$
                                                       &&$<0.227$
                                                       &$<0.558$
                                                       &&$<0.877$
                                                       &$0.400^{+0.150}_{-0.350}$\\

%$m_{{\rm sterile}}^{\rm thermal}\,[\rm eV]$       &$...$
%                                                  &$...$
 %                                                 &&$...$
%                                                  &$...$
 %                                                 &&$...$
 %                                                 &$1.635$\\
%\hline
%$\chi^2_{\rm min}$&13663.816&13706.838&&13664.796&13709.380&&13664.690&13706.908\\
%\hline
\hline
\end{tabular}

\end{table*}

%%%%%%%%%%%%%%%%%%%%%%%%%%%%%%%%%%%%%%%%%%%%%%%%%%%%%%%%%%%%%%%%

%%%table%%%%%planck+bsh%%%%%%%%%%%%%%%%%%%%%%%%%%%%%%%%%%%%%%%%%%%%%%%%%%%%%%%%%%%%
\begin{table*}\small
\setlength\tabcolsep{1.0pt}
\renewcommand{\arraystretch}{1.5}
\centering
\caption{\label{tabnovs}Fit results for the IvCDM (with fixed neutrino mass $\sum m_{\nu}=0.06$ eV), IvCDM +$\nu_s$, and IvCDM+$\nu_a$ models from the data combination Planck+BSH.}
\begin{tabular}{ccccccccc}

\hline

  &\multicolumn{3}{c}{$Q=\beta H\rho_{\rm v}$}&&\multicolumn{3}{c}{$Q=\beta H\rho_{\rm c}$}\\
           \cline{2-4}\cline{6-8}
  Model &IvCDM1&IvCDM1+$\nu_s$&IvCDM1+$\nu_a$&&IvCDM2&IvCDM2+$\nu_s$&IvCDM2+$\nu_a$\\
\hline
$\Omega_m$
                            &$0.321^{+0.061}_{-0.053}$
                            &$0.333^{+0.069}_{-0.053}$  %%%IvCDM1+$\nu_s$
                            &$0.316^{+0.067}_{-0.049}$
                            &&$0.294\pm0.007$
                            &$0.296\pm0.008$             %%%IvCDM2+$\nu_s$
                            &$0.295^{+0.007}_{-0.008}$\\

$\sigma_8$
                            &$0.821^{+0.036}_{-0.032}$
                            &$0.818^{+0.040}_{-0.039}$  %%%IvCDM1+$\nu_s$
                            &$0.827\pm0.035$
                            &&$0.845^{+0.016}_{-0.015}$
                            &$0.821^{+0.037}_{-0.024}$  %%%IvCDM2+$\nu_s$
                            &$0.844^{+0.019}_{-0.017}$\\

$H_0\,[{\rm km}/{\rm s}/{\rm Mpc}]$
                                                 &$68.09\pm0.48$
                                                 &$69.50\pm1.10$           %%%IvCDM1+$\nu_s$
                                                 &$68.19^{+0.55}_{-0.49}$
                                                 &&$68.94^{+0.61}_{-0.60}$
                                                 &$69.50^{+0.70}_{-0.92}$   %%%IvCDM2+$\nu_s$
                                                 &$68.93\pm0.64$\\
$\beta$
                          &$-0.08^{+0.22}_{-0.27}$
                          &$-0.13^{+0.22}_{-0.30}$  %%%IvCDM1+$\nu_s$
                          &$-0.04^{+0.20}_{-0.30}$
                          &&$0.0021\pm0.0011$
                          &$0.0022\pm0.0015$        %%%IvCDM2+$\nu_s$
                          &$0.0021^{+0.0012}_{-0.0013}$\\

%$\sum m_{\nu}$            &...
%                          &...
%                          &$<0.10$
%                          &...
%                          &...
%                          &$<0.20$\\

\hline
\end{tabular}

\end{table*}
%%%%%%%%%%%%%%%%%%%%%%%%%%%%%%%%%%%%%%%%%%%%%%%%%%%%%%%%%%%%%%%%%%%%%%%%%%%%%%%%%%%%%%%%%%%%%%%%%%%%%%%%5
%%%table%%%planck+bsh+lss%%%%%%%%%%%%%%%%%%%%%%%%%%%%%%%%%%%%%%%%%%%%%%%%%%%%%%%%%%%%%%
\begin{table*}\small
\setlength\tabcolsep{1.0pt}
\renewcommand{\arraystretch}{1.5}
\centering
\caption{\label{tabnovs2}Fit results for the IvCDM (with fixed neutrino mass $\sum m_{\nu}=0.06$ eV), IvCDM +$\nu_s$, and IvCDM+$\nu_a$ models from the data combination Planck+BSH+LSS.}
\begin{tabular}{ccccccccc}

\hline

  &\multicolumn{3}{c}{$Q=\beta H\rho_{\rm v}$}&&\multicolumn{3}{c}{$Q=\beta H\rho_{\rm c}$}\\
           \cline{2-4}\cline{6-8}
  Model &IvCDM1&IvCDM1+$\nu_s$&IvCDM1+$\nu_a$&&IvCDM2&IvCDM2+$\nu_s$&IvCDM2+$\nu_a$\\
\hline
$\Omega_m$               &$0.316^{+0.033}_{-0.030}$
                         &$0.310^{+0.039}_{-0.031}$
                         &$0.315^{+0.033}_{-0.029}$
                         &&$0.297\pm0.008$
                         &$0.297\pm0.008$
                         &$0.298\pm0.008$\\

$\sigma_8$               &$0.805\pm0.018$
                         &$0.793\pm0.021$
                         &$0.804\pm0.019$
                         &&$0.819\pm0.011$
                         &$0.785^{+0.026}_{-0.022}$
                         &$0.813^{+0.014}_{-0.013}$\\

$H_0\,[{\rm km}/{\rm s}/{\rm Mpc}]$           &$68.23\pm0.47$
                                              &$68.84^{+0.70}_{-1.09}$
                                              &$68.19^{+0.60}_{-0.52}$
                                              &&$68.73\pm0.62$
                                              &$69.31\pm0.70$
                                              &$68.57\pm0.66$\\

$\beta$                 &$-0.06^{+0.13}_{-0.15}$
                        &$-0.02^{+0.14}_{-0.18}$
                        &$-0.05^{+0.12}_{-0.15}$
                        &&$0.0014\pm0.0011$
                        &$0.0026\pm0.0015$
                        &$0.0021^{+0.0014}_{-0.0016}$\\

\hline
\end{tabular}

\end{table*}
%%%%%%%%%%%%%%%%%%%%%%%%%%%%%%%%%%%%%%%%%%%%%%%%%%%%%%%%%%%%%%%%%%%%%%%%%%%%%%%%%%%%%%%%%%%%%%%%%%%%%%%%5

%%%figde%%%%%%%%%%%%%%%%%
\begin{figure*}[!htp]
\includegraphics[scale=0.25]{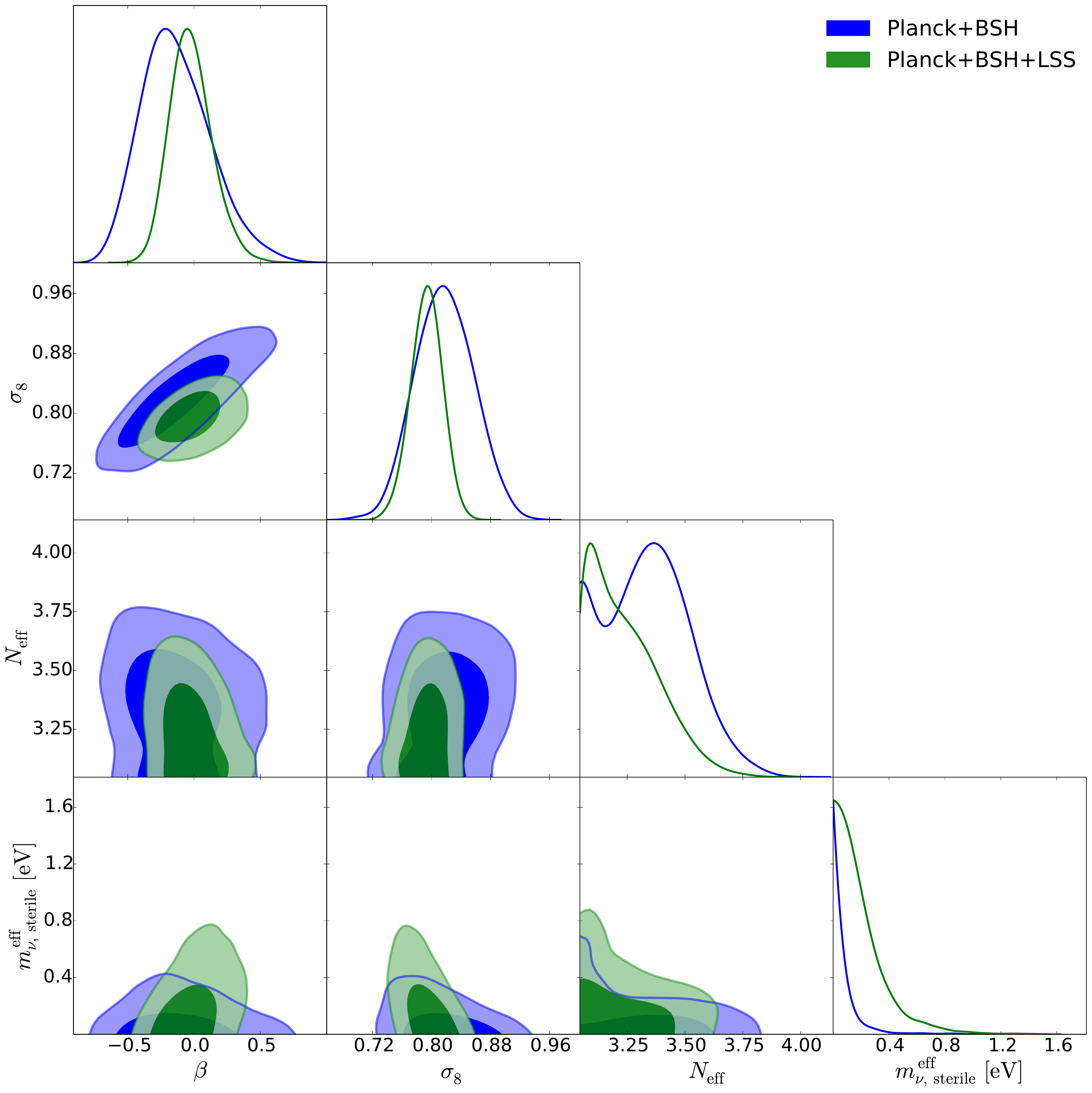}
\centering
\caption{\label{devs}The one-dimensional posterior distributions and two-dimensional marginalized contours (1$\sigma$ and 2$\sigma$) for parameters $\beta$, $\sigma_8$, $N_{\rm eff}$, and $m_{\nu,{\rm sterile}}^{\rm eff}$ of the I$\Lambda$CDM1 ($Q=\beta H\rho_{\rm v}$)+$\nu_s$ model by using the Planck+BSH ({\it blue}) and the Planck+BSH+LSS ({\it green}) data combinations.}
\end{figure*}

%%%%%%%%%%%%%%%%%%%%%%%%%

%%%figc%%%%%%%%%%%%%%%%%
\begin{figure*}[!htp]
\includegraphics[scale=0.25]{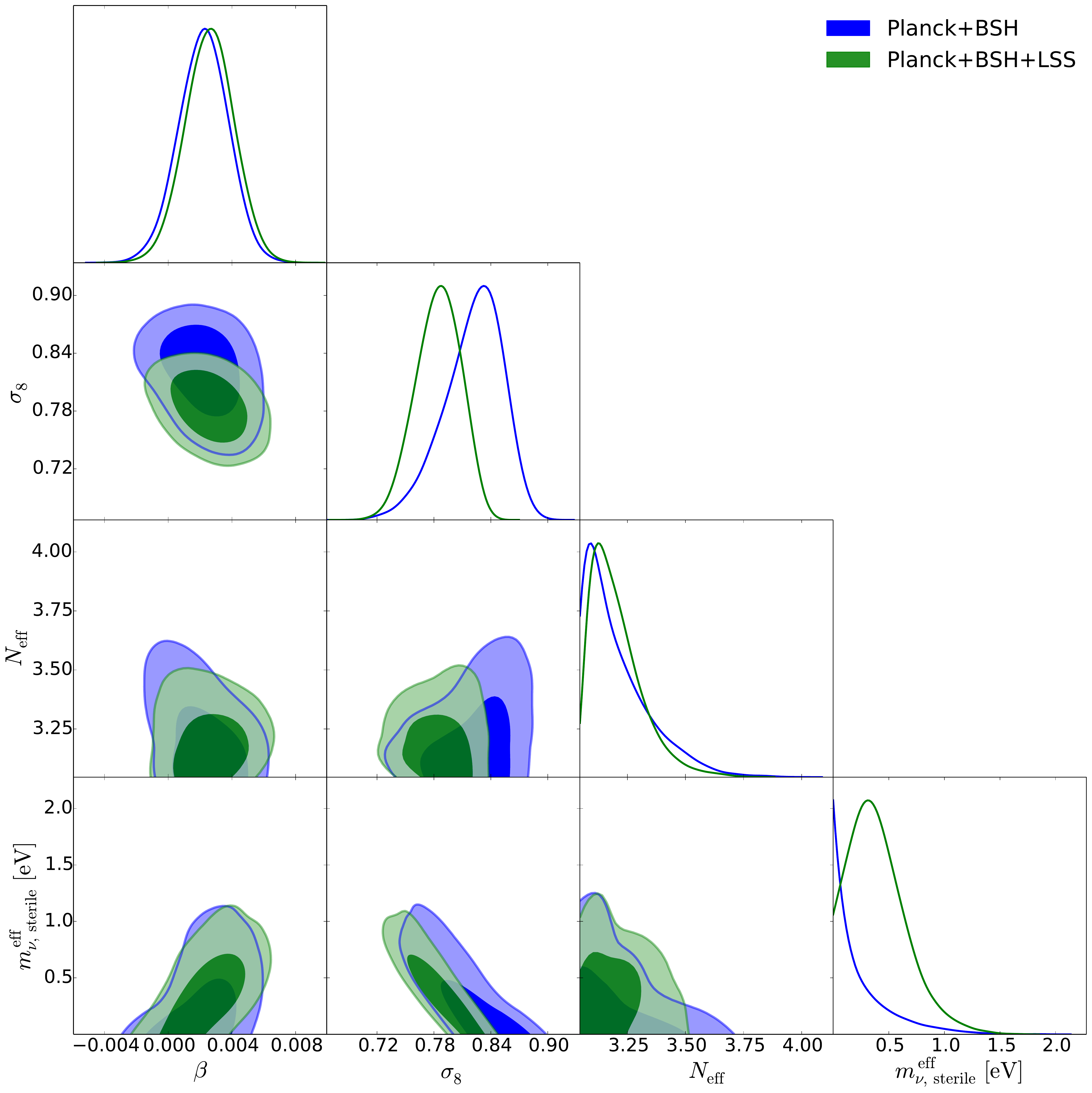}
\centering
\caption{\label{cvs}The one-dimensional posterior distributions and two-dimensional marginalized contours (1$\sigma$ and 2$\sigma$) for parameters  $\beta$, $\sigma_8$, $N_{\rm eff}$, and $m_{\nu,{\rm sterile}}^{\rm eff}$ of the IvCDM2 ($Q=\beta H\rho_c$)+$\nu_s$ model by using the Planck+BSH ({\it blue}) and the Planck+BSH+LSS ({\it green}) data combinations.}
\end{figure*}

%%%fig:allvs%%%%%%%%%%%%%%%%%
\begin{figure*}[!htp]
\includegraphics[scale=0.35]{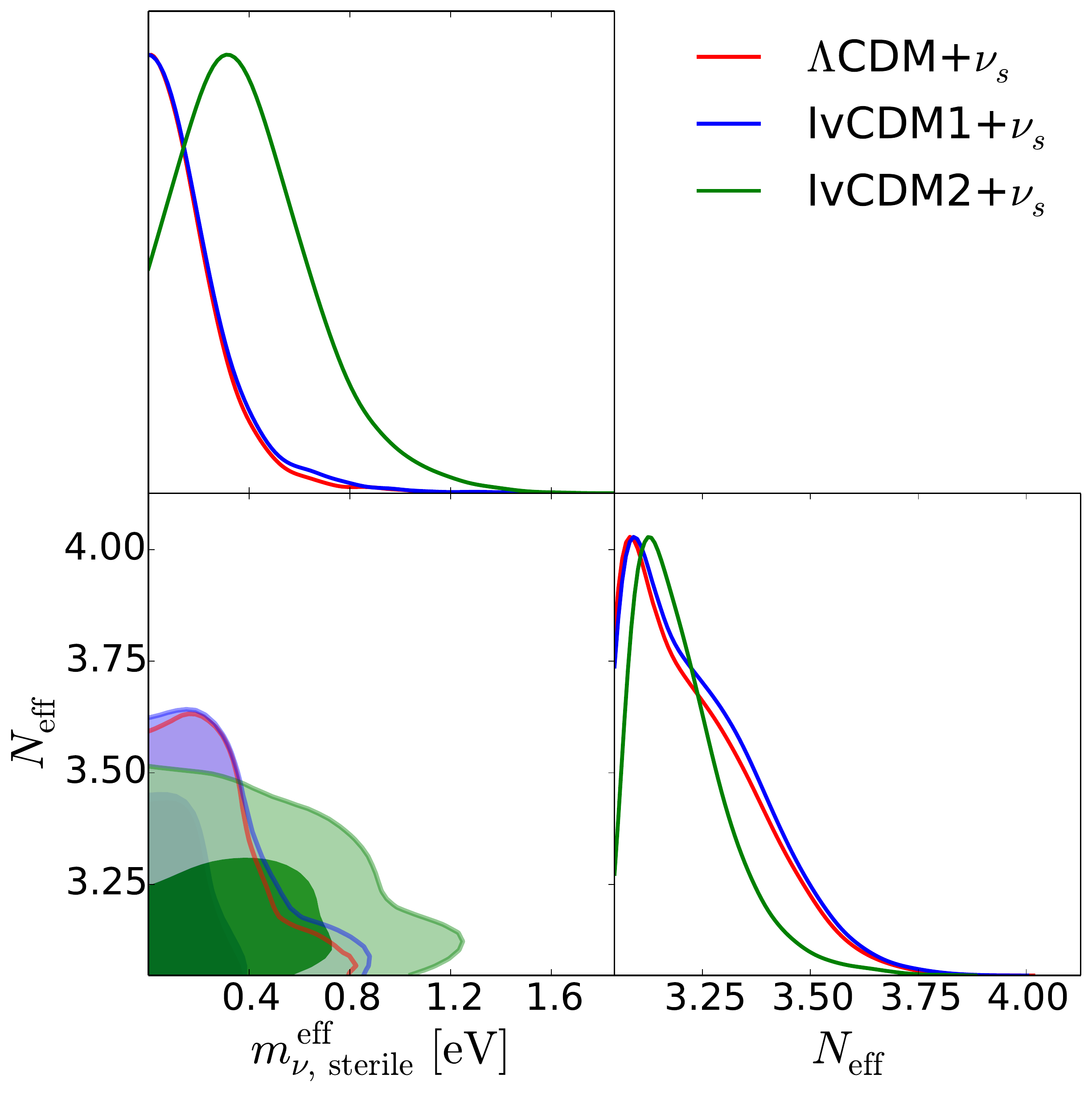}
\centering
\caption{\label{allvs}The one-dimensional posterior distributions and two-dimensional marginalized contours (1$\sigma$ and 2$\sigma$) for parameters $m_{\nu,{\rm sterile}}^{\rm eff}$ and $N_{\rm eff}$ of the $\Lambda$CDM+$\nu_s$, IvCDM1 ($Q=\beta H\rho_{\rm v}$)+$\nu_s$, and IvCDM2 ($Q=\beta H\rho_{\rm c}$)+$\nu_s$ models by using the Planck+BSH+LSS data combination.}
\end{figure*}
%%%%%%%%%%%%%%%%%%%%%%%%%

%%%figvs+va%%%%%%%%%%%%%%%%%
\begin{figure*}[!htp]
\includegraphics[scale=0.20]{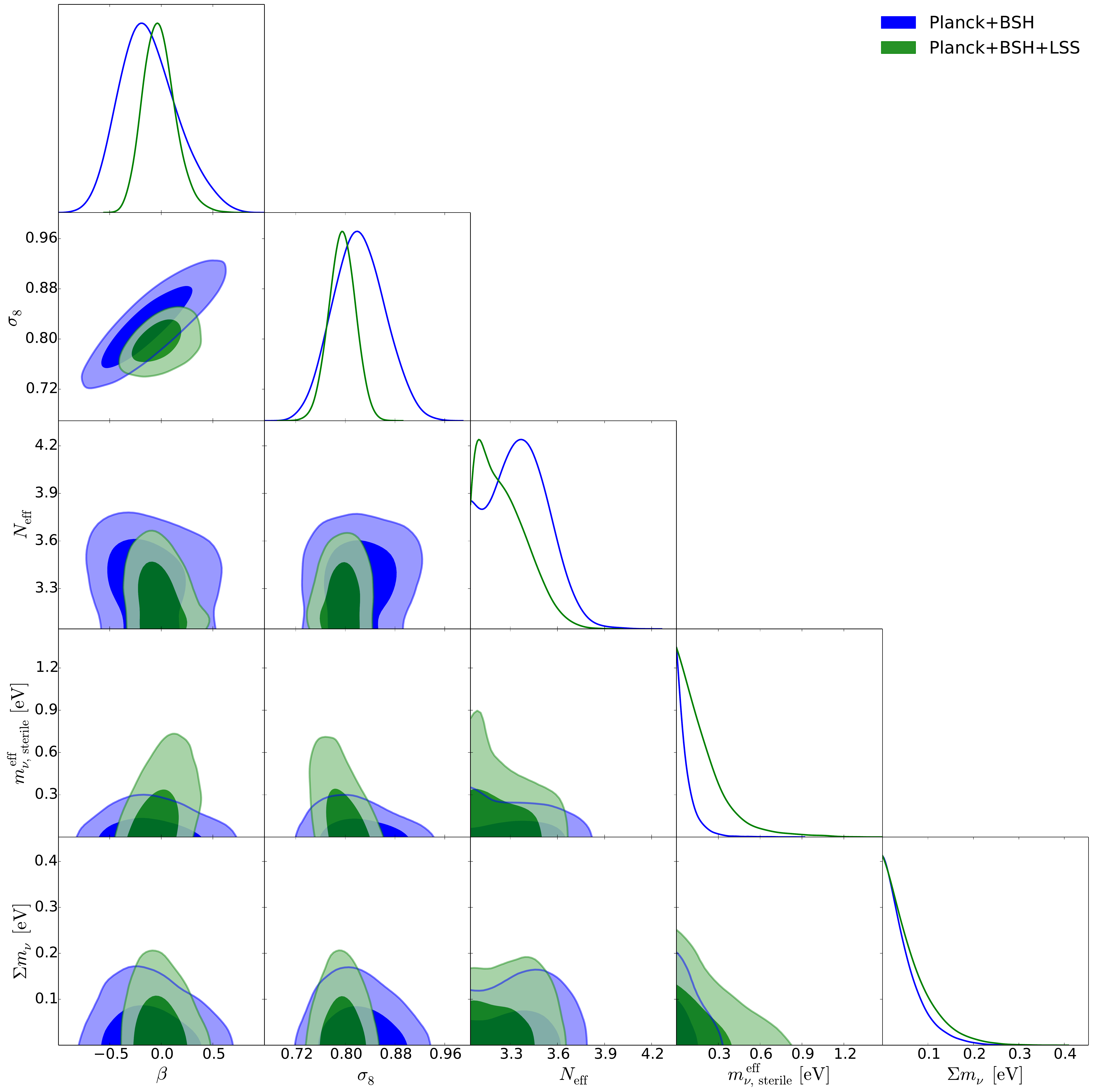}
\centering
\caption{\label{devsa}The one-dimensional posterior distributions and two-dimensional marginalized contours (1$\sigma$ and 2$\sigma$) for parameters $\beta$, $\sigma_8$, $N_{\rm eff}$, $m_{\nu,{\rm sterile}}^{\rm eff}$, and $\sum m_\nu$ of the IvCDM1 ($Q=\beta H\rho_{\rm de}$)+$\nu_s$+$\nu_a$ model by using the Planck+BSH ({\it blue}) and Planck+BSH+LSS ({\it green}) data combinations.}
\end{figure*}
%%%%%%%%%%%%%%%%%%%%%%%%%%%%%%%%%%%%%%%%%%%%%%%%%%%%%%%%%%%%%%%%%%%%%%%%%%%%%%%%%
%%%figvs+va%%%%%%%%%%%%%%%%%
\begin{figure*}[!htp]
\includegraphics[scale=0.20]{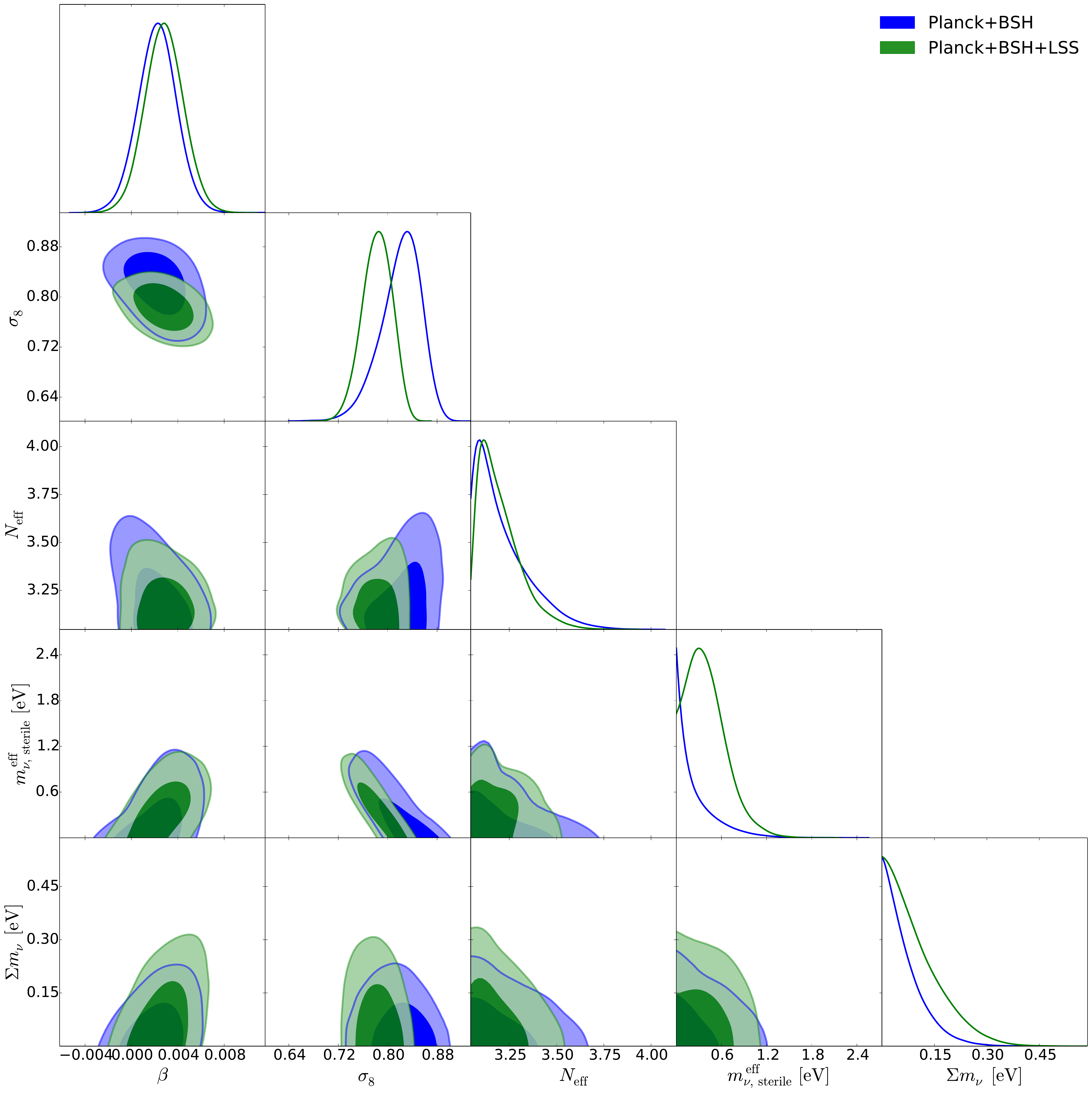}
\centering
\caption{\label{cvsa}The one-dimensional posterior distributions and two-dimensional marginalized contours (1$\sigma$ and 2$\sigma$) for parameters $\beta$, $\sigma_8$, $N_{\rm eff}$, $m_{\nu,{\rm sterile}}^{\rm eff}$, and $\sum m_\nu$ of the IvCDM2 ($Q=\beta H\rho_{\rm c}$)+$\nu_s$+$\nu_a$ model by using the Planck+BSH ({\it blue}) and Planck+BSH+LSS ({\it green}) data combinations.}
\end{figure*}
%%%%%%%%%%%%%%%%%%%%%%%%%%%%%%%%%%%%%%%%%%%%%%%%%%%%%%%%%%%%%%%%%%%%%%%%%%%%%%%%%

%%%figde:beta%%%%%%%%%%%%%%%%%
\begin{figure*}[!htp]
\includegraphics[scale=0.65]{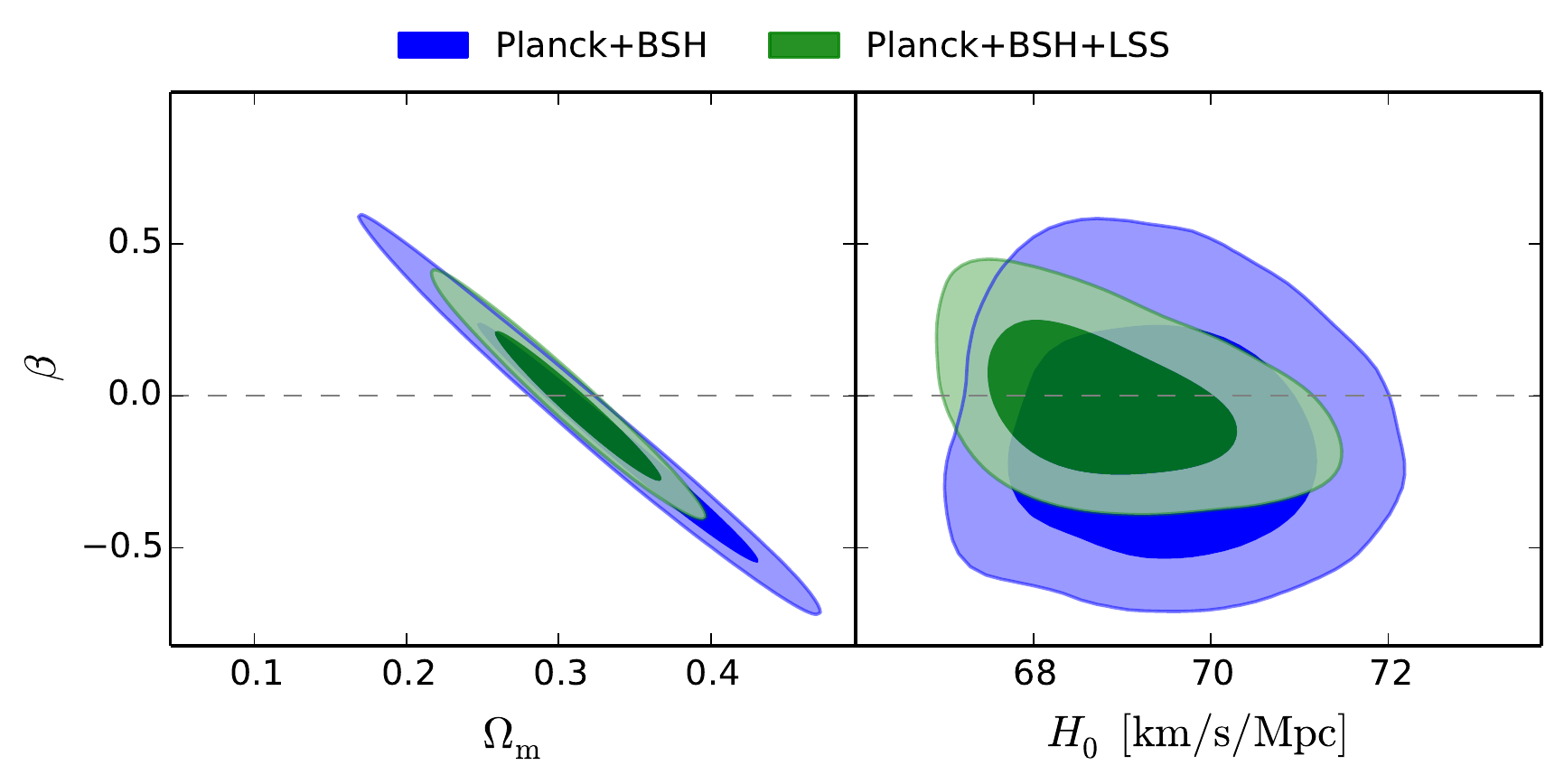}
\centering
\caption{\label{figde}The two-dimensional marginalized contours (1$\sigma$ and 2$\sigma$) in the $\Omega_m$-$\beta$ and $H_0$-$\beta$ planes for the IvCDM1 ($Q=\beta H\rho_{\rm v}$)+$\nu_s$ model from the data combinations of Planck+BSH and Planck+BSH+LSS.}
\end{figure*}
%%%%%%%%%%%%%%%%%%%%%%%%%
%%%figc:beta%%%%%%%%%%%%%%%%%
\begin{figure*}[!htp]
\includegraphics[scale=0.65]{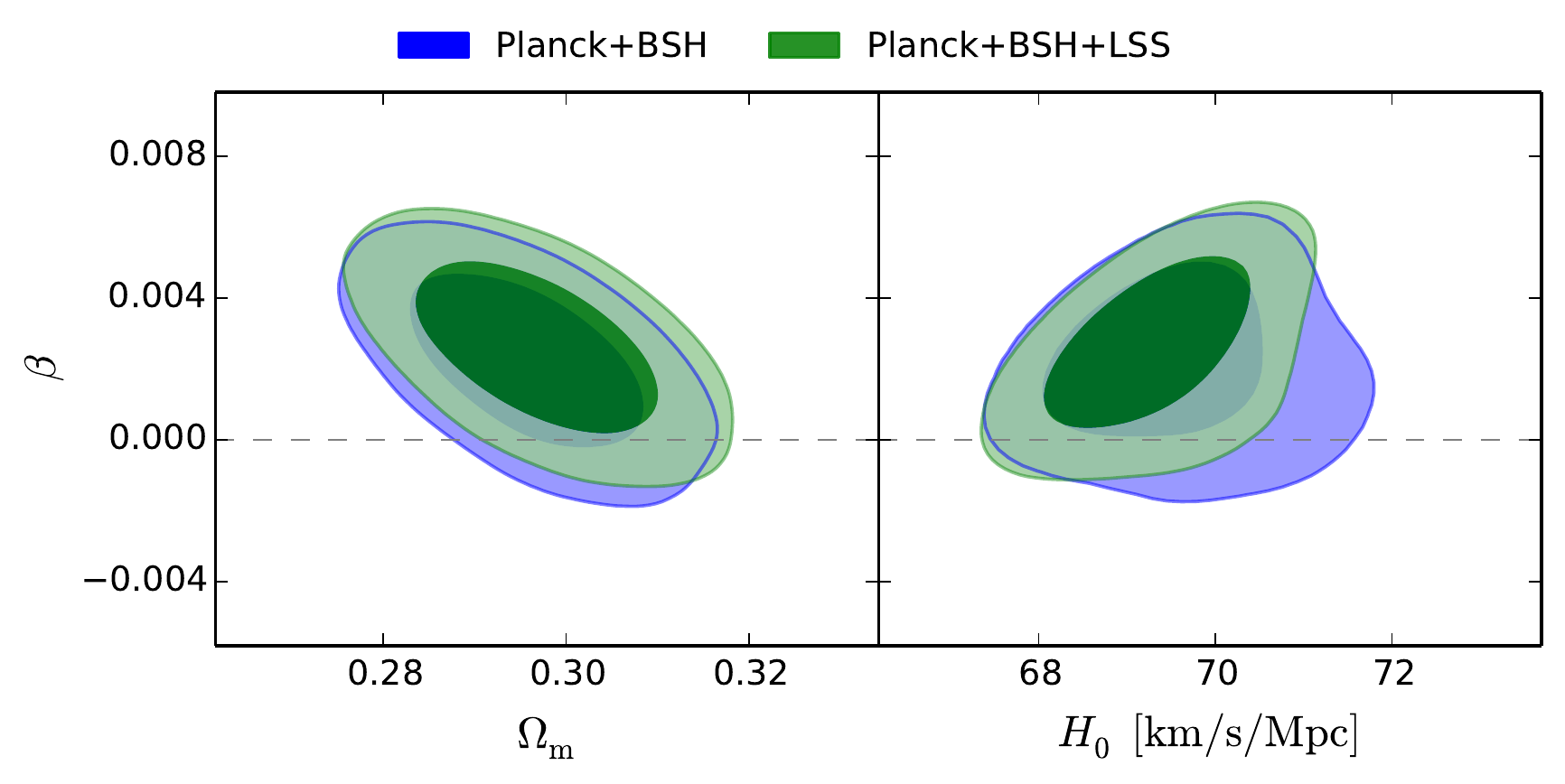}
\centering
\caption{\label{figc}The two-dimensional marginalized contours (1$\sigma$ and 2$\sigma$) in the $\Omega_m$-$\beta$ and $H_0$-$\beta$ planes for the IvCDM2 ($Q=\beta H\rho_{\rm c}$)+$\nu_s$ model from the data combinations of Planck+BSH and Planck+BSH+LSS.}
\end{figure*}
%%%%%%%%%%%%%%%%%%%%%%%%%%%%%%%%%%%%%%
%%%figlike+cbsh%%%%%%%%%%%%%%%%%
\begin{figure*}[!htp]
\includegraphics[scale=0.7]{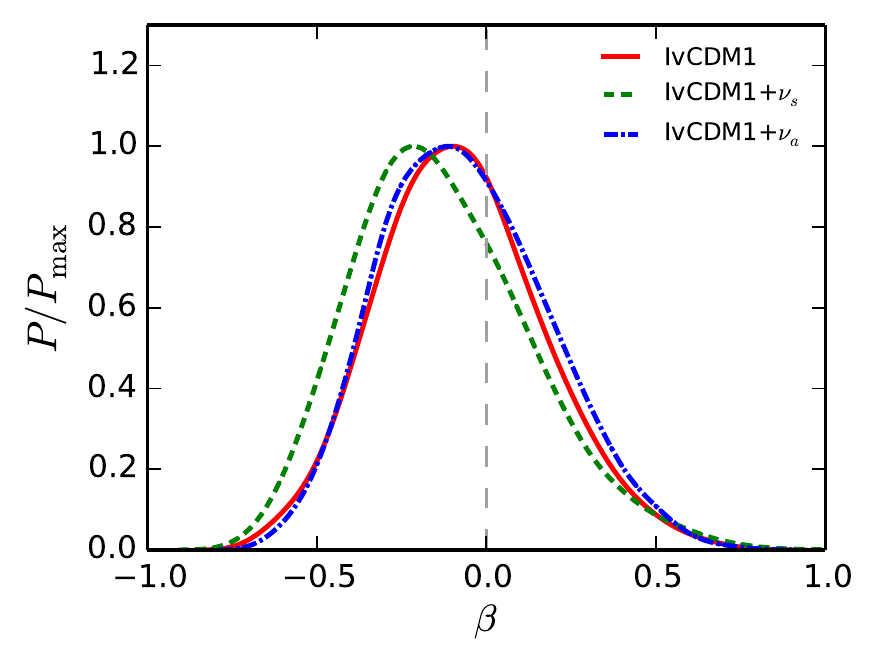}
\includegraphics[scale=0.7]{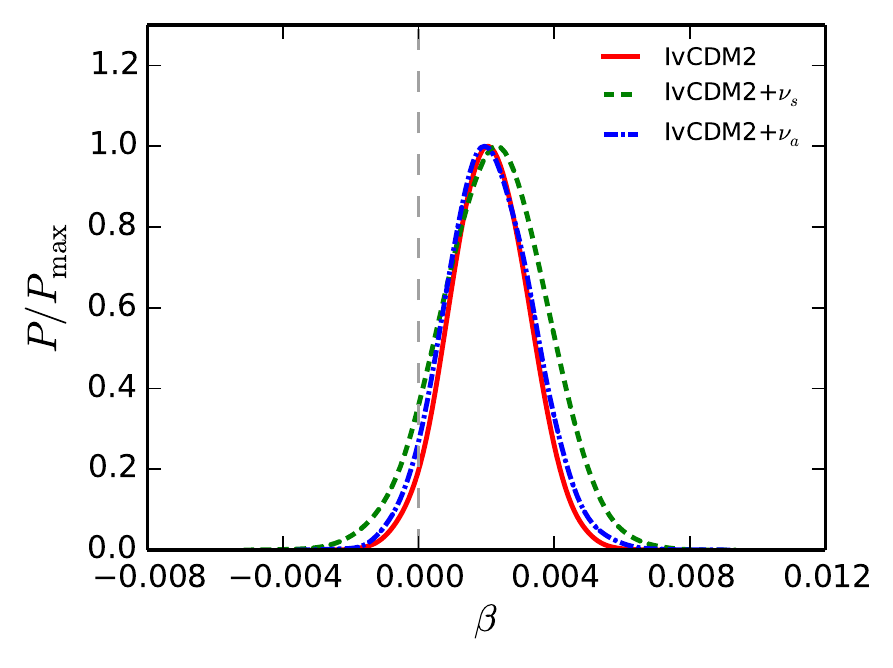}
\centering
\caption{\label{figlike1}The one-dimensional posterior distributions for the coupling parameter $\beta$ by using Planck+BSH. The left panel corresponds to the IvCDM1, IvCDM1+$\nu_s$, and IvCDM1+$\nu_a$ models, and the right panel corresponds to the IvCDM2, IvCDM2+$\nu_s$, and IvCDM2+$\nu_a$ models.}
\end{figure*}
%%%%%%%%%%%%%%%%%%%%%%%%%
%%%figlike+cbshwrl%%%%%%%%%%%%%%%%%
\begin{figure*}[!htp]
\includegraphics[scale=0.7]{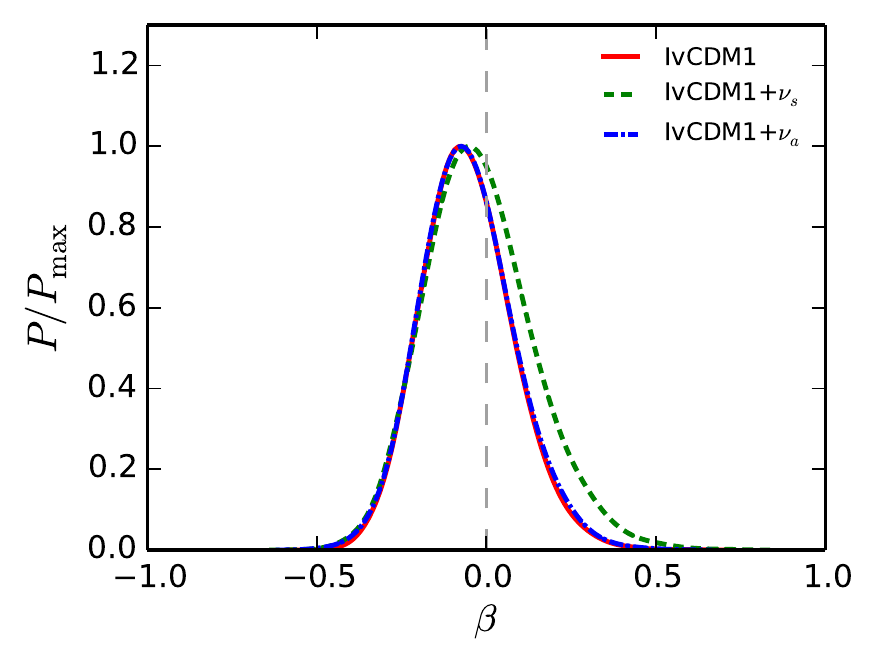}
\includegraphics[scale=0.7]{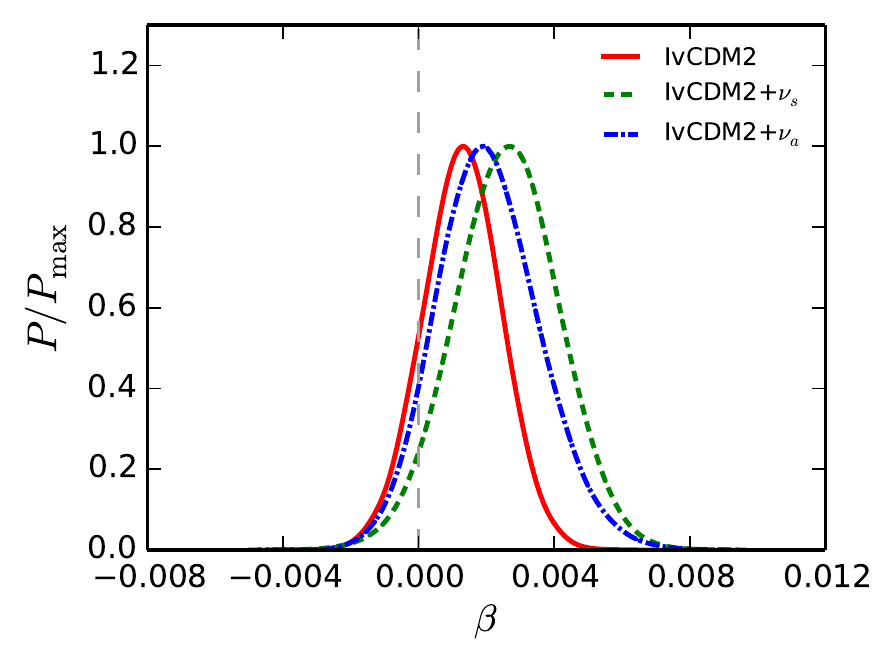}
\centering
\caption{\label{figlike2}The one-dimensional posterior distributions for the coupling parameter $\beta$ by using Planck+BSH+LSS. The left panel corresponds to the IvCDM1, IvCDM1+$\nu_s$, and IvCDM1+$\nu_a$ models, and the right panel corresponds to the IvCDM2, IvCDM2+$\nu_s$, and IvCDM2+$\nu_a$ models.}
\end{figure*}

%%%%%%%%%%%%%%%%%%%%%%%%%

%%%figh0like%%%%%%%%%%%%%%%%%
\begin{figure*}[!htp]
\includegraphics[scale=0.7]{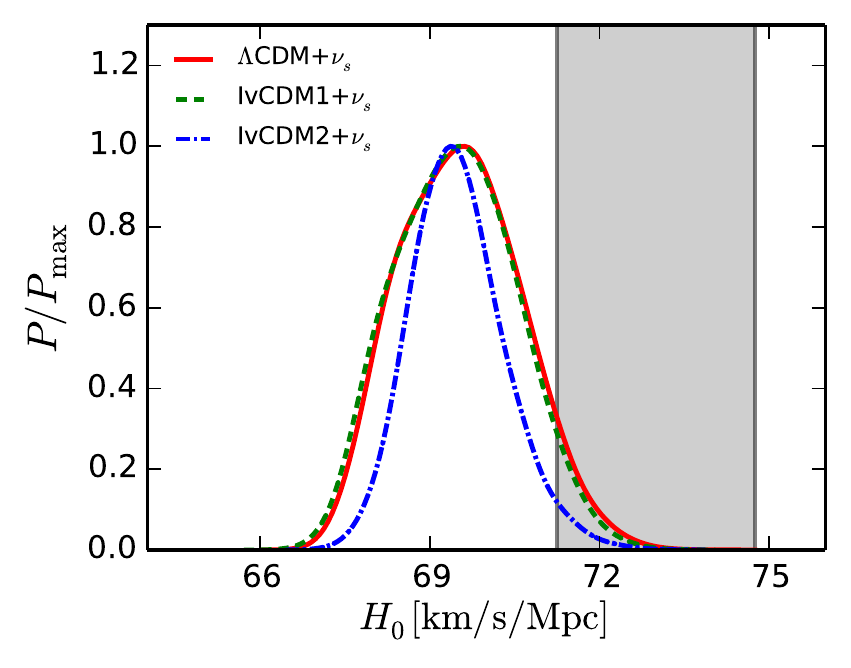}
\includegraphics[scale=0.7]{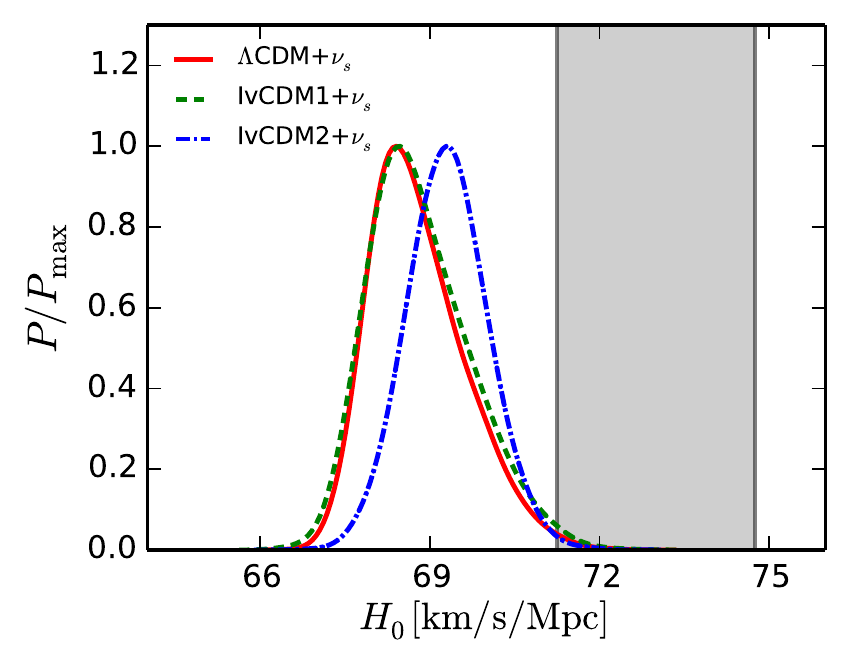}
\centering
\caption{\label{figlikeh0}The one-dimensional posterior distributions of $H_0$ for the $\Lambda$CDM+$\nu_s$, IvCDM1+$\nu_s$, and IvCDM2+$\nu_s$ models from the constraints of the Planck+BSH ({\it left}) and Planck+BSH+LSS ({\it right}) data combinations. The result of direct measurement of the Hubble constant is shown by a grey band.}
\end{figure*}
%%%%%%%%%%%%%%%%%%%%%%%%%

\section{Results and discussion}\label{sec3}

%In this section, we investigate the constraint results on the sterile neutrino parameters $N_{\rm eff}$ and $m_{\nu,{\rm sterile}}^{\rm eff}$ and coupling constant $\beta$ in a universe of vacuum energy interacting with cold dark matter.

Detailed fit results for cosmological parameters are given in Tables~\ref{tabvs}--\ref{tabnovs2} and Figs.~\ref{devs}--\ref{figlikeh0}. In the tables, we quote $\pm1\sigma$ errors, but for the parameters that cannot be well constrained, we only give $2\sigma$ upper limits.
%For simplicity, in the cases of active neutrinos, we use IvCDM1+$\nu_a$ and IvCDM2+$\nu_a$ to denote the corresponding $Q=\beta H\rho_{\rm v}$ and $Q=\beta H\rho_{\rm c}$ models. In the cases of consider both active and sterile neutrinos in the models, we use $\Lambda$CDM+$\nu_s$+$\nu_a$, IvCDM1+$\nu_s$+$\nu_a$, and IvCDM2+$\nu_s$+$\nu_a$ to denote the corresponding $Q=0$, $Q=\beta H\rho_{\rm v}$, and $Q=\beta H\rho_{\rm c}$ models.

\subsection{Sterile neutrino parameters}

In this subsection, we present the constraint results of the IvCDM1 ($Q=\beta H\rho_{\rm v}$)+$\nu_s$ and IvCDM2 ($Q=\beta H\rho_{\rm c}$)+$\nu_s$ models obtained by using Planck+BSH and Planck+BSH+LSS data combinations and analyze how the interaction between vacuum energy and cold dark matter affects the cosmological constraints on the sterile neutrino parameters. We take the $\Lambda$CDM+$\nu_s$ model as a reference model in this study, and thus the fitting results of this model by using the same data combinations are also given.
Besides, we also make a simple discussion for the models in which both active and sterile neutrinos are considered.
The main constraint results are given in Tables~\ref{tabvs} and \ref{tabvsva} and Figs.~\ref{devs}--\ref{cvsa}.

For the IvCDM1+$\nu_s$ model, we obtain $N_{\rm eff}<3.641$ and $m_{\nu,{\rm sterile}}^{\rm eff}<0.312$ eV from Planck+BSH, and $N_{\rm eff}<3.522$ and $m_{\nu,{\rm sterile}}^{\rm eff}<0.576$ eV from Planck+BSH+LSS.
We find that these constraint results are basically consistent with those of the $\Lambda$CDM+$\nu_s$ model by using both two data combinations.
We notice that in this model adding the LSS data leads to a larger effective sterile neutrino mass limit. This is because the current LSS observations favor a lower matter perturbation (demonstrated by a lower $\sigma_8$), and in this case $m_{\nu,{\rm sterile}}^{\rm eff}$ is anti-correlated with $\sigma_8$ due to the free-streaming of sterile neutrinos (see Fig.~\ref{devs}), thus a larger effective sterile neutrino mass limit is derived. This result is in accordance with the conclusions in previous studies for the active neutrino mass~\cite{Guo:2017hea}.
In addition, obviously, we see that only upper limits on $N_{\rm eff}$ and $m_{\nu,{\rm sterile}}^{\rm eff}$ are obtained by using both two data combinations, which means that in this case only weak hint of the existence of massive sterile neutrinos is found.

For the IvCDM2+$\nu_s$ model, we obtain $N_{\rm eff}<3.498$ and $m_{\nu,{\rm sterile}}^{\rm eff}<0.875$ eV from Planck+BSH. We find that, compared with the $\Lambda$CDM+$\nu_s$ model, the constraint on $N_{\rm eff}$ becomes slightly tighter, but the constraint on $m_{\nu,{\rm sterile}}^{\rm eff}$ becomes much looser. Figure~\ref{cvs} shows that $\beta$ is positive correlated with $m_{\nu,{\rm sterile}}^{\rm eff}$. Thus a larger $\beta$ will lead to a larger $m_{\nu,{\rm sterile}}^{\rm eff}$, which explains why the limit of $m_{\nu,{\rm sterile}}^{\rm eff}$ derived in the IvCDM2+$\nu_s$ model is much larger than that in the $\Lambda$CDM+$\nu_s$ model. This result is also in accordance with the conclusions in previous studies for the active neutrino mass~\cite{Guo:2017hea}, i.e., it was found in Ref.~\cite{Guo:2017hea} that the IvCDM2+$\nu_a$ model leads to a much looser limit on the total mass of active neutrinos ($\sum m_\nu < 0.20$~eV) with the same data combination Planck+BSH. Similarly, in the case of the sterile neutrino species, the IvCDM2+$\nu_s$ model also leads to a much looser upper limit on the effective mass of sterile neutrino ($m_{\nu,{\rm sterile}}^{\rm eff}<0.875$ eV).
In addition, the Planck+BSH+LSS constraint gives $N_{\rm eff}=3.204^{+0.049}_{-0.135}$ and $m_{\nu,{\rm sterile}}^{\rm eff}=0.410^{+0.150}_{-0.330}$ eV.
Evidently, adding the measurements of growth of structure tightens the constraints on $N_{\rm eff}$ and $m_{\nu,{\rm sterile}}^{\rm eff}$ significantly.
Thus, our results prefer $\Delta N_{\rm eff}>0$ at the 1.17$\sigma$ level and a non-zero mass at the 1.24$\sigma$ level. In this case, according to our fit results, we have $\Delta N_{\rm eff}\approx0.16$ and $m_{{\rm sterile}}^{\rm thermal}\approx 1.636$ eV, indicating a partially thermalized sterile neutrino with eV-scale mass.

To directly show how the forms of interaction affect the constraints on sterile neutrino parameters, we show in Fig.~\ref{allvs} the comparison of the three models, i.e., $\Lambda$CDM+$\nu_s$, IvCDM1 ($Q=\beta H\rho_{\rm v}$)+$\nu_s$, and IvCDM2 ($Q=\beta H\rho_{\rm c}$)+$\nu_s$, from Planck+BSH+LSS combination (tighter constraint).
We can clearly see that, in the IvCDM1 ($Q=\beta H\rho_{\rm v}$)+$\nu_s$ model, the posterior distribution curves and the marginalized contours in the $m_{\nu,{\rm sterile}}^{\rm eff}$--$N_{\rm eff}$ plane are almost in coincidence with the $\Lambda$CDM+$\nu_s$ model. The constraint results for parameters of the IvCDM2+$\nu_s$ model are evidently different.

Furthermore, we also consider both active and sterile neutrinos in the interacting vacuum energy models. The main constraint results are shown in Table~\ref{tabvsva} and Figs.~\ref{devsa}~and~\ref{cvsa}.

From Table~\ref{tabvsva}, we find that in these models (the $\Lambda$CDM+$\nu_s$+$\nu_a$, IvCDM1+$\nu_s$+$\nu_a$, and IvCDM2+$\nu_s$+$\nu_a$ models), the fit values of parameters ($\beta$, $N_{\rm eff}$, $m_{\nu,{\rm sterile}}^{\rm eff}$, and so on) are actually similar to those in the models without considering active neutrinos (see Table~\ref{tabvs}).
To show apparently the effect of active neutrinos on the constraints on parameters in IvCDM+$\nu_s$+$\nu_a$ models,
the one- and two-dimensional marginalized posterior distributions of parameters ($\beta$, $\sigma_8$,  $N_{\rm eff}$, $m_{\nu,{\rm sterile}}^{\rm eff}$, and $\sum m_\nu$) for the IvCDM1+$\nu_s$+$\nu_a$ and IvCDM2+$\nu_s$+$\nu_a$ models are plotted in Figs.~\ref{devsa}~and~\ref{cvsa}, respectively. We can clearly see that, compared with Figs.~\ref{devs}~and~\ref{cvs}, the one- and two-dimensional marginalized posterior distributions are very similar.
Therefore, the consideration of active neutrinos in the models with sterile neutrinos almost does not influence fitting results and also does not introduce new degeneracies.

\subsection{Coupling constant $\beta$}

In this subsection, we present the results of coupling constant $\beta$. The main constraint results are shown in Tables~\ref{tabvs}--\ref{tabnovs2} and Figs.~\ref{figde}--\ref{figlike2}.

For the IvCDM1+$\nu_s$ model, from Table~\ref{tabvs}, we have $\beta=-0.13^{+0.22}_{-0.30}$ by using Planck+BSH, and $\beta=-0.02^{+0.14}_{-0.18}$ by using Planck+BSH+LSS. These results slightly favor a negative coupling constant, which means  that vacuum energy decays into cold dark matter.
For this model, the two-dimensional marginalized posterior distribution contours including $\beta$ are shown in Fig.~\ref{figde}. One can clearly see that in this model $\beta=0$ is actually well consistent with the current observations inside the 1$\sigma$ range, and further including the LSS data (WL, RSD, and CMB lensing) can lead to tighter constraint results for the parameters. Moreover, for both data combinations, we find that there exists strong degeneracy between $\beta$ and $\Omega_m$ in this model.

For the IvCDM2+$\nu_s$ model, we obtain $\beta=0.0022\pm0.0015$ from Planck+BSH, and $\beta=0.0026\pm0.0015$ from Planck+BSH+LSS.
A remarkable feature of this model is that the coupling constant $\beta$ can be tightly constrained for the two data combinations.
Another interesting phenomenon is that the both two data combinations consistently favor a positive coupling constant $\beta$ at more than 1$\sigma$ level, indicating that the case of cold dark matter decaying into vacuum energy is more supported. The results of $\beta>0$ (at more than 1$\sigma$ level) can be clearly seen in Fig.~\ref{figc}.

Here, with the purpose of directly showing the effect of sterile neutrinos on the constraints of the coupling constant $\beta$, we also perform an analysis for the IvCDM model (without sterile neutrinos), to make a comparison, and the detailed fit results are given in Tables~\ref{tabnovs} and \ref{tabnovs2} as well as Figs.~\ref{figlike1} and \ref{figlike2}.

In Tables~\ref{tabnovs}~and~\ref{tabnovs2}, for a direct comparison, we duplicate the constraint results of the IvCDM+$\nu_s$ models from the Planck+BSH and Planck+BSH+LSS data combinations here (see also Table~\ref{tabvs}). By using the Planck+BSH data, we obtain $\beta=-0.08^{+0.22}_{-0.27}$ for the IvCDM1 model and $\beta=0.0021\pm0.0011$ for the IvCDM2 model. In the cases with sterile neutrinos, we have $\beta=-0.13^{+0.22}_{-0.30}$ for the IvCDM1+$\nu_s$ model and $\beta=0.0022\pm0.0015$ for the IvCDM2+$\nu_s$ model.
By using the Planck+BSH+LSS data, we obtain $\beta=-0.06^{+0.13}_{-0.15}$ for the IvCDM1 model and $\beta=0.0014\pm0.0011$ for the IvCDM2 model. When sterile neutrinos are considered in these models, we have $\beta=-0.02^{+0.14}_{-0.18}$ for the IvCDM1+$\nu_s$ model and $\beta=0.0026\pm0.0015$ for the IvCDM2+$\nu_s$ model.
One can clearly see that, in the IvCDM1 cases (with $Q=\beta H\rho_{\rm v}$), $\beta=0$ is inside the 1$\sigma$ range, no matter if the massive sterile neutrino is taken into account. On the other hand, in the IvCDM2 cases (with $Q=\beta H\rho_{\rm c}$), both the IvCDM2 model and the IvCDM2+$\nu_s$ model favor $\beta>0$ at more than  1$\sigma$ level.
That is to say, the consideration of sterile neutrinos in the IvCDM scenario almost does not influence the constraint results of the coupling constant $\beta$.

We also wish to compare our analysis with previous studies for the active neutrino mass in the interacting vacuum energy model~\cite{Guo:2017hea}. In Ref.~\cite{Guo:2017hea} (see also Table~\ref{tabnovs}), it was found that $\beta=-0.04^{+0.20}_{-0.30}$ for the IvCDM1+$\nu_a$ model and $\beta=0.0021^{+0.0012}_{-0.0013}$ for the IvCDM2+$\nu_a$ model by using the Planck+BSH data combination. In the case of using the Planck+BSH+LSS data combination, we have $\beta=-0.05^{+0.12}_{-0.15}$ for the IvCDM1+$\nu_a$ model and $\beta=0.0021^{+0.0014}_{-0.0016}$ for the IvCDM2+$\nu_a$ model.
Obviously, we find that the consideration of massive (sterile/active) neutrinos almost does not influence the constraints on $\beta$ in both the $Q=\beta H\rho_{\rm v}$ model and the $Q=\beta H\rho_{\rm c}$ model.
The one-dimensional posterior distributions of $\beta$ for the three models (IvCDM, IvCDM+$\nu_s$, and IvCDM+$\nu_a$) using Planck+BSH and Planck+BSH+LSS are shown in Figs.~\ref{figlike1}~and~\ref{figlike2}.

\subsection{The tension issue}
In this subsection, we wish to simply address the issue of the $H_0$ tension between the Planck observation and the direct measurement from distance ladder. The tension has been discussed extensively in the literature, by additionally considering massive active neutrinos, sterile neutrinos, or dynamical dark energy. In this subsection, we only focus on the interaction between vacuum energy and cold dark matter, and we wish to see whether the interaction can play a significant role in relieving the tension.
Our discussion is based on the constraint results shown in Table~\ref{tabvs} (see also Fig.~\ref{figlikeh0}).

From Table~\ref{tabvs}, we see that in the case of using the Planck+BSH data combination, the $\Lambda$CDM+$\nu_s$ model gives $H_0=69.50^{+1.00}_{-1.20}~{\rm km}~{\rm s}^{-1}~{\rm Mpc}^{-1}$, the IvCDM1+$\nu_s$ model gives $H_0=69.50{\pm1.10}~{\rm km}~{\rm s}^{-1}~{\rm Mpc}^{-1}$, and the IvCDM2+$\nu_s$ model gives $H_0=69.50^{+0.70}_{-0.92}~{\rm km}~{\rm s}^{-1}~{\rm Mpc}^{-1}$, indicating that the tensions with the direct measurement of the Hubble constant are at the 1.74$\sigma$, 1.69$\sigma$, and 1.86$\sigma$ levels, respectively. We thus find that in the IvCDM1+$\nu_s$ model, the tension is only slightly relieved compared with the $\Lambda$CDM+$\nu_s$ model, while in the IvCDM2+$\nu_s$ model the tension is even slightly enhanced.
In the case of using the Planck+BSH+LSS data combination, we obtain $H_0=68.80^{+0.61}_{-1.03}~{\rm km}~{\rm s}^{-1}~{\rm Mpc}^{-1}$ in the $\Lambda$CDM+$\nu_s$ model, $H_0=68.84^{+0.70}_{-1.09}~{\rm km}~{\rm s}^{-1}~{\rm Mpc}^{-1}$ in the IvCDM1+$\nu_s$ model, and $H_0=69.31{\pm0.70}~{\rm km}~{\rm s}^{-1}~{\rm Mpc}^{-1}$ in the IvCDM2+$\nu_s$ model. So, in this case, the tensions with the direct measurement  are at the 2.27$\sigma$, 2.21$\sigma$, and 1.96$\sigma$ levels, respectively, showing that the consideration of the interaction between vacuum energy and cold dark matter of this form can only slightly relieve the tension.

Therefore, from the analysis above, we can clearly see that the consideration of interaction between vacuum energy and cold dark matter only offers a marginal improvement for the $H_0$ tension in the most cases.
To directly show how the interaction affects the constraints on $H_0$, we plot one-dimensional posterior distributions for $H_0$ in Fig.~\ref{figlikeh0}.

\section{Conclusion}\label{sec4}

In this paper, we consider massive sterile neutrinos in the interacting vacuum energy scenario. We have studied two interacting vacuum energy models with the energy transfer rates $Q=\beta H\rho_{\rm v}$ and $Q=\beta H\rho_{\rm c}$, respectively.
We used the PPF approach to calculate the perturbations of the vacuum energy in the interacting scenario.
In this paper, we have two aims:
(i) We wish to investigate if the interaction (between vacuum energy and cold dark matter) and sterile neutrino parameters have some correlations with each other.
(ii) We wish to search for sterile neutrinos in the interacting vacuum energy model with the latest cosmological observations.
The observational data used in this paper include the Planck 2015 TT,TE,EE+lowP data, the BAO measurements, the SN data, the latest direct measurement of $H_0$, the shear data of WL observation, the RSD measurement, and the Planck lensing measurement.

We show that, in the IvCDM model with $Q=\beta H\rho_{\rm v}$, $N_{\rm eff}<3.641$ and $m_{\nu,{\rm sterile}}^{\rm eff}<0.312$ eV are obtained from Planck+BSH, and $N_{\rm eff}<3.522$ and $m_{\nu,{\rm sterile}}^{\rm eff}<0.576$ eV are obtained from Planck+BSH+LSS.
In the IvCDM model with $Q=\beta H\rho_{\rm c}$, we obtain $N_{\rm eff}<3.498$ and $m_{\nu,{\rm sterile}}^{\rm eff}<0.875$ eV from Planck+BSH, and $N_{\rm eff}=3.204^{+0.049}_{-0.135}$ and $m_{\nu,{\rm sterile}}^{\rm eff}=0.410^{+0.150}_{-0.330}$ eV from Planck+BSH+LSS.
Thus, for the IvCDM model with $Q=\beta H\rho_{\rm v}$, only upper limits on $N_{\rm eff}$ and $m_{\nu,{\rm sterile}}^{\rm eff}$ are obtained by using both two data combinations. For the IvCDM model with $Q=\beta H\rho_{\rm c}$, using the Planck+BSH data, only upper limits on $N_{\rm eff}$ and $m_{\nu,{\rm sterile}}^{\rm eff}$ can be derived. Further including the LSS (WL+RSD+lensing) data significantly improves the constraints, and we have $N_{\rm eff}=3.204^{+0.049}_{-0.135}$ and $m_{\nu,{\rm sterile}}^{\rm eff}=0.410^{+0.150}_{-0.330}$ eV using Planck+BSH+LSS. According to the fit results of the latter case, we have $\Delta N_{\rm eff}\approx0.16$ and $m_{{\rm sterile}}^{\rm thermal}\approx 1.636$ eV, indicating a partially thermalized sterile neutrino with eV-scale mass.
Therefore, we conclude that the different interacting forms could influence the constraint results of sterile neutrino parameters significantly.
%Besides, when we consider the active neutrinos into IvCDM+$\nu_s$, we find that the introduction of active neutrinos has little effect on fitting results and degeneracy for the IvCDM+$\nu_s$+$\nu_a$ model.

According to the constraints of two data combinations, we find that for the IvCDM1+$\nu_s$ model $\beta=0$ is consistent with the current data inside 1$\sigma$ range, which implies that this model is recovered to a standard $\Lambda$CDM universe. For the IvCDM2+$\nu_s$ model, we find that $\beta>0$ is favored at more than 1$\sigma$ level, which indicates that cold dark matter decays into vacuum energy, and we show that this model can be tightly constrained by using the two data combinations.
We find that the consideration of massive sterile neutrinos almost does not influence the constraint on the coupling constant $\beta$.

%In addition, we also address the $H_0$ tension issue. We find that, in some cases the consideration of the interaction between vacuum energy and dark matter offers a marginal improvement for the $H_0$ tension.

\begin{acknowledgments}
This work was supported by the National Natural Science Foundation of China (Grants Nos.~11835009, 11875102, 11522540, and 11690021), the Top-Notch Young Talents Program of China, and the Provincial Department of Education of Liaoning (Grant No.~L2012087).

\end{acknowledgments}

\end{document}